\documentclass[%
 reprint,
superscriptaddress,
 amsmath,amssymb,
 aps,
]{revtex4-2}

\usepackage{graphicx}
\usepackage{dcolumn}
\usepackage{bm}
\usepackage{siunitx}
 \usepackage{relsize}
\usepackage{placeins}
\usepackage{subfig}
\usepackage{caption}
\usepackage{times}
\usepackage{xcolor} 
\usepackage{hyperref}
\hypersetup{
    colorlinks=true,
    linkcolor=blue,
citecolor=blue}

\makeatletter
\let\save@mathaccent\mathaccent
\newcommand*\if@single[3]{%
  \setbox0\hbox{${\mathaccent"0362{#1}}^H$}%
  \setbox2\hbox{${\mathaccent"0362{\kern0pt#1}}^H$}%
  \ifdim\ht0=\ht2 #3\else #2\fi
  }
\newcommand*\rel@kern[1]{\kern#1\dimexpr\macc@kerna}
\newcommand*\widebar[1]{\@ifnextchar^{{\wide@bar{#1}{0}}}{\wide@bar{#1}{1}}}
\newcommand*\wide@bar[2]{\if@single{#1}{\wide@bar@{#1}{#2}{1}}{\wide@bar@{#1}{#2}{2}}}
\newcommand*\wide@bar@[3]{%
  \begingroup
  \def\mathaccent##1##2{%
    \let\mathaccent\save@mathaccent
    \if#32 \let\macc@nucleus\first@char \fi
    \setbox\z@\hbox{$\macc@style{\macc@nucleus}_{}$}%
    \setbox\tw@\hbox{$\macc@style{\macc@nucleus}{}_{}$}%
    \dimen@\wd\tw@
    \advance\dimen@-\wd\z@
    \divide\dimen@ 3
    \@tempdima\wd\tw@
    \advance\@tempdima-\scriptspace
    \divide\@tempdima 10
    \advance\dimen@-\@tempdima
    \ifdim\dimen@>\z@ \dimen@0pt\fi
    \rel@kern{0.6}\kern-\dimen@
    \if#31
      \overline{\rel@kern{-0.6}\kern\dimen@\macc@nucleus\rel@kern{0.4}\kern\dimen@}%
      \advance\dimen@0.4\dimexpr\macc@kerna
      \let\final@kern#2%
      \ifdim\dimen@<\z@ \let\final@kern1\fi
      \if\final@kern1 \kern-\dimen@\fi
    \else
      \overline{\rel@kern{-0.6}\kern\dimen@#1}%
    \fi
  }%
  \macc@depth\@ne
  \let\math@bgroup\@empty \let\math@egroup\macc@set@skewchar
  \mathsurround\z@ \frozen@everymath{\mathgroup\macc@group\relax}%
  \macc@set@skewchar\relax
  \let\mathaccentV\macc@nested@a
  \if#31
    \macc@nested@a\relax111{#1}%
  \else
    \def\gobble@till@marker##1\endmarker{}%
    \futurelet\first@char\gobble@till@marker#1\endmarker
    \ifcat\noexpand\first@char A\else
      \def\first@char{}%
    \fi
    \macc@nested@a\relax111{\first@char}%
  \fi
  \endgroup
}
\makeatother

\newsavebox{\mybox}
\newlength{\mywidth}
\newlength{\myheight}
\newlength{\myline}
\newlength{\myoffset}
\newcommand{\mysqrt}[1]%
{\setlength{\myline}{.1ex}%
\addtolength{\myline}{.06pt}%
\setlength{\myoffset}{.9em}
\addtolength{\myoffset}{-2pt}
\savebox{\mybox}{$\displaystyle\sqrt{#1}$}%
\settoheight{\myheight}{\usebox{\mybox}}%
\addtolength{\myheight}{-.2ex}
\settowidth{\mywidth}{\usebox{\mybox}}%
\addtolength{\mywidth}{-\myoffset}%
 \rlap{\usebox{\mybox}}\hspace{\myoffset}{\raisebox{\myheight}{\rule{\mywidth}{\myline}}}}

\begin{document}

\preprint{APS/123-QED}

\title{Diffuseness parameter as a bottleneck for accurate half-life calculations}

\author{Aladdin Abdul-latif}
\affiliation{Department of Physics, German University in Cairo, Cairo 11835, Egypt}
\affiliation{Department of Physics, Cairo University, Giza 12613, Egypt}
\author{Omar Nagib}%
\email{{\color{black} omar.khaled.nagib@gmail.com}}
\affiliation{Department of Physics, German University in Cairo, Cairo 11835, Egypt}

\date{\today}

\begin{abstract}
 An investigation of the calculated $\alpha$ decay half-lives of super heavy nuclei (SHN) reveals that the diffuseness parameter is a great bottleneck for achieving accurate results and predictions. In particular, when universal proximity function is adopted for nuclear potential, half-life is found to vary significantly and nonlinearly as a function of diffuseness parameter. To overcome this limiting hurdle, a new semiempirical formula for diffuseness that is dependent on charge and neutron numbers is proposed in this work. With the model at hand, half-lives of 218 SHN are computed, for 68 of which there exists available experimental data and 150 of which are predicted. The calculations of half-lives for 68 SHN are compared against experimental data and the calculated data obtained by using deformed Woods-Saxon, deformed Coulomb potentials model, and six semiempirical formulas. The predictions of 150 SHN are compared against the predictions of seven of the current best semiempirical formulas. Calculations of the present study are in good agreement with the experimental half-lives outperforming all but ImSahu semiempirical formula. Moreover, the predictions of our model are consistent with predictions of the semiempirical formulas. We strongly conclude that more attention should be directed toward obtaining accurate diffuseness parameter values for using it in nuclear calculations. 
\end{abstract}

\maketitle


\section{\label{sec:level1}Introduction}

Currently, within the theoretical framework of nuclear physics, especially in $\alpha$ decay of super heavy nuclei (SHN), there are points of intersection and divergence between different models used in calculations. Particularly, points of disagreement abound, including the particular functional form of nuclear interaction between $\alpha$/cluster and the daughter nuclei, be it given by Woods-Saxon (WS) potential \cite{probe}, double folding model (DFM) \cite{DFM1,DFM2}, liquid drop model (LDM) \cite{LDM}, universal proximity potential \cite{prox,proxies} and et cetera. Another point of disagreement stems from two competing proposals for $\alpha$ decay. The first--often dubbed cluster-like theory-- assigns a preformation factor $P_o$ for every parent nuclei, while the second fission-like model asserts that $P_o=1$ for all nuclei \cite{freq}. On the other hand, the diffuseness parameter $a$ is one of the parameters with universal presence in all models regardless of the particular nuclear potential under study. Although ubiquitous, little attention is paid to the parameter which is often set constant for all nuclei \cite{probe,consta1, consta2, consta3, consta4}. More recently, Dehghani \emph{et al.} investigated the role diffuseness parameter plays when deformed Woods-Saxon potential is adopted for nuclear interaction between $\alpha$ and daughter nuclei. Some interesting conclusions were reached of which we list the chief ones \cite{roleconst}: 
\begin{enumerate}
  \item The logarithm of half-life $\log_{10} T_{1/2}$ decreases linearly with increasing diffuseness parameter. For instance, a variation of 0.4 fm of $a$ can induce a change in the logarithm of half-life by 2 (i.e., half-life changes by two orders of magnitude). 
  \item  For a given nucleus, a systematic search was carried out in search for the best $a$ value that matches the experimental half-life for that particular nucleus. This process was applied on 68 SHN and diffuseness parameter for 68 SHN was extracted out of experimental data of half-lives. 
  \item Adopting a constant $a=0.54$ fm for all nuclei was found to be optimal in the sense of minimizing the root-mean-square (rms) error of the logarithm of half-lives. The rms error came out to be 0.787, implying that on average, adopting $a=0.54$ fm will induce an error of 0.787 for the logarithm of half-life. 
\end{enumerate}
Our work extends and generalizes these results by investigating the effect of diffuseness parameter when universal proximity potential is adopted for nuclear interaction between $\alpha$ and daughter nuclei. The outline of this paper is as follows: in Sec. \ref{sec:num2}, we outline the theoretical framework that we will be working with regarding calculations of half-lives. In Sec. \ref{sec:num3}, the relation between diffuseness and half-life is probed when proximity potential is adopted and results are compared with the case of deformed WS, deformed Coulomb potentials. In Sec. \ref{sec:num4}, we investigate the relation between diffuseness parameter on one hand and charge, neutron and mass numbers on the other, and we propose a new semiempirical formula for diffuseness as a function of charge and neutron numbers. In Sec. \ref{sec:num5}, calculations and predictions of 218 SHN are made; calculations for 68 SHN will be carried out using proximity potential with variable effective diffuseness and compared with calculations of deformed WS, deformed Coulomb potentials in which diffuseness parameter is set to $a=0.54$ fm. Moreover, our calculations will be compared with six popular semiempirical formulas for half-lives. The viability of the new formula for diffuseness is tested by predicting the half-lives of 150 SHN and comparing the results with seven of the current best semiempirical formulas. We conclude the last section by discussing the most important highlights, results, and potential research directions in the future.

\section{Theoretical framework} \label{sec:num2}
Generally, the effective potential for $\alpha$ decay can be broken into three contributions, namely nuclear, Coulomb, and angular contributions

\begin{equation} \label{eq:1}
V_{\text{eff}}(r)= V_N(r)+V_C(r)+V_l(r)
\end{equation}

where $r$ is the separation distance between the center of the daughter and $\alpha$ nuclei. For our particular potential, we assume spherical symmetry and ignore deformation effects, i.e., we consider the potential to assume radial dependence only, with the quadrupole and hexadecapole deformation parameters $\beta_2$ and $\beta_4$ set to zero. Moreover, we shall consider cases in which there's no total angular momentum carried by the $\alpha$-daughter system and hence $V_l(r)$ is subsequently ignored. Regarding the nuclear term $V_N(r)$, we adopt the universal proximity potential previously proposed by Zhang \emph{et al.} \cite{prox}:

\begin{equation} \label{eq:2}
V_N(s_0)=4 \pi b_{\text{eff}} \widebar{R} \gamma \phi(s_0)
\end{equation}

\begin{equation} \label{eq:3}
 \phi(s_0)=\dfrac{p_1}{1+\exp \Big(\dfrac{s_0+p_2}{p_3}\Big)}
\end{equation}

In two parameter Fermi distribution (2pF) of nuclear matter, $b_{\text{eff}}$ and the effective diffuseness $a_{\text{eff}}$ of the proximity potential are related as

\begin{equation} \label{eq:4}
b_{\text{eff}} =\dfrac{\pi}{\sqrt{3}} a_{\text{eff}}
\end{equation}

The effective diffuseness is a function of both the nuclear surface diffuseness of the daughter and $\alpha$ nuclei. Following previous works \cite{prox,Z94}, we posit the ansatz that $a_{\text{eff}}$ is the average of the two, i.e.,  $a_{\text{eff}}=(a_{\alpha}+a_d)/2$ with  $a_{\alpha}=0.3238$ fm and $a_d$ refers to the nuclear surface diffuseness of the daughter. We shall explain and show hereafter that adopting a constant $a_{\text{eff}}$ for all nuclei leads to unacceptable errors, especially for the proximity potential and for systems in which we expect that  $a_{\text{eff}} \ge 0.5$ fm (or equivalently $b_{\text{eff}} \ge 0.9$ fm). $\gamma$ is given by

\begin{equation} \label{eq:5}
\gamma =0.9517 \Big [1-1.7826\Big(\dfrac{N-Z}{A}\Big)^2 \Big] \,   \si{\mega\eV \per  \femto \metre \squared}
\end{equation}

$ \widebar{R}=R_{\alpha}R_d/(R_{\alpha}+R_d)$ is the reduced radius of the $\alpha$-daughter system where the radius of each nucleus in terms of its mass number is given by  

\begin{equation} \label{eq:6}
R = 1.28 A^{1/3}+0.8A^{-1/3}-0.76
\end{equation}

$ \phi(s_0)$ is the universal function that quantifies the nuclear interaction between $\alpha$ and the daughter in terms of the reduced separation distance $s_0$

\begin{equation} \label{eq:7}
s_0=\dfrac{r-R_{\alpha}-R_d}{b_{\text{eff}}}
\end{equation}

In particular, this equation holds in the regime $s_0>-1$. The constants $p_1,p_2$ and $p_3$ appearing in Eq. \eqref{eq:3} are given by -7.65, 1.02 and 0.89 respectively. Moving on, the Coulomb term $V_C(r)$ is given by

\begin{equation}\label{eq:8}
      V_C(r)=
      Z_{\alpha} Z_d e^2 \left\{ \begin{array}{ll}
            \dfrac{1}{r} & r\geq R_C \\
            \dfrac{1}{2 R_C}\Big[3-\Big(\dfrac{r}{R_C}\Big)^2\Big] & r< R_C
        \end{array} \right.
    \end{equation}

where  $Z_{\alpha}$ and $Z_d$ are the $\alpha$ and daughter charge numbers and $R_C=R_{\alpha} +R_{d}$. In the context of WKB approximation, the half-life is given by

\begin{equation} \label{eq:9}
T_{1/2}=\dfrac{\pi \hbar \ln2}{P_o E_{\nu}}\Big(1+\exp(K) \Big)
\end{equation}

where $P_o,  E_{\nu}$ and $K$ are the preformation factor, zero-point vibration energy and action integral respectively. The action integral is given by

\begin{equation} \label{eq:10}
K= \dfrac{2}{\hbar} \int_{r_1}^{r_2} \sqrt{2 \mu (V_{\text{eff}}(r)-Q_{\alpha} )} dr
\end{equation}

where $Q_{\alpha}, \mu, r_1$, and $r_2$ are the decay energy, reduced mass, and first and second turning points, respectively. We shall describe $E_{\nu}$ classically, since it was previously found \cite{freq} that classical and quantum mechanical (e.g., modified harmonic oscillator) approaches yield not too different results, hence

\begin{equation} \label{eq:11}
E_{\nu}=\dfrac{\hbar \omega}{2}=\dfrac{\hbar \pi}{2 R_p}\sqrt{\dfrac{2 E_{\alpha}}{m_{\alpha}}}
\end{equation}

where $R_p,E_{\alpha}$ and $m_{\alpha}$ are the parent radius, kinetic energy of $\alpha$ nuclei and its mass respectively. The decay energy of the system and the kinetic energy of $\alpha$ nuclei are related by the recoil of the daughter and electron shielding, or in other words \cite{extended},

\begin{equation} \label{eq:12}
E_{\alpha}=\dfrac{A_d}{A_p} Q_{\alpha}- \Big(6.53 Z_d^{7/5}-8 Z_d^{2/5} \Big)\cdot 10^{-5} \,  \si{\mega\eV}
\end{equation}

where $A_d$ and $A_p$ are the mass numbers of the daughter and the parent nuclei respectively. Finally, for the preformation factor, we adopt \cite{preform} 

\begin{multline} \label{eq:13}
 \log_{10} P_o = a+b(Z-Z_1)(Z_2-Z_1)+c(N-N_1)(N_2-N) \\
 +d A+e(Z-Z_1)(N-N_1)
\end{multline}

where $a,b,c,d$ and $e$ are given by 34.90593, 0.003011, 0.003717, -0.151216 and 0.006681 respectively. For $82 \le Z \le 126$ and $152 \le N  \le 184$, which is the scope of the present paper, the magic numbers $Z_1,Z_2,N_1$ and $N_2$ are given by 82, 126, 152 and 184 respectively.

\section{Diffuseness and half-life for proximity potential} \label{sec:num3}

\emph{A priori}, one should not expect the effective diffuseness parameter $a_{\text{eff}}$ for the proximity potential and the diffuseness parameter for WS potential $a_{\text{\tiny WS}}$ to be equal or even play the same logical roles; this is owing to the different functional forms of the two potentials. To be more precise, consider the undeformed WS potential \cite{probe} 

\begin{equation} \label{eq:14}
V_{\text{\tiny WS}}(r)=\dfrac{-V_0}{1+\exp \Big(\dfrac{r-R_0}{a_{\text{\tiny WS}}}\Big)}
\end{equation}

where $V_0$ is the depth of the potential well and $R_0$ is the half radius at which $V(R_0)= -V_0/2$. One can easily see that letting $a_{\text{\tiny WS}} \rightarrow \infty$ leads to $V_{\text{\tiny WS}}(r) \rightarrow -V_0/2$ for all $r$, while $a_{\text{\tiny WS}} \rightarrow 0$ leads to  $V_{\text{\tiny WS}}(r) \rightarrow -V_0$ for $r \le R_0$. In other words, smaller $a_{\text{\tiny WS}}$ leads to a deeper potential well (and vice versa) with the depth of the potential varying from $-V_0$ to $-V_0/2$.

Regarding the proximity potential, however, from Eqs. \eqref{eq:2}, \eqref{eq:3} and \eqref{eq:7}, we see that $a_{\text{eff}} \rightarrow \infty$ leads to $V_N(r) \rightarrow -\infty$ for all $r$, while $a_{\text{eff}} \rightarrow 0$ leads to  $V_N(r) \rightarrow 0$ for all $r$; that is, greater  $a_{\text{eff}}$ leads to a deeper potential well (unlike the WS case) and vice versa, with the depth of the potential varying from $-\infty$ to 0.

\begin{figure}[htp!]
\centering
\includegraphics[width=0.5\textwidth,keepaspectratio]{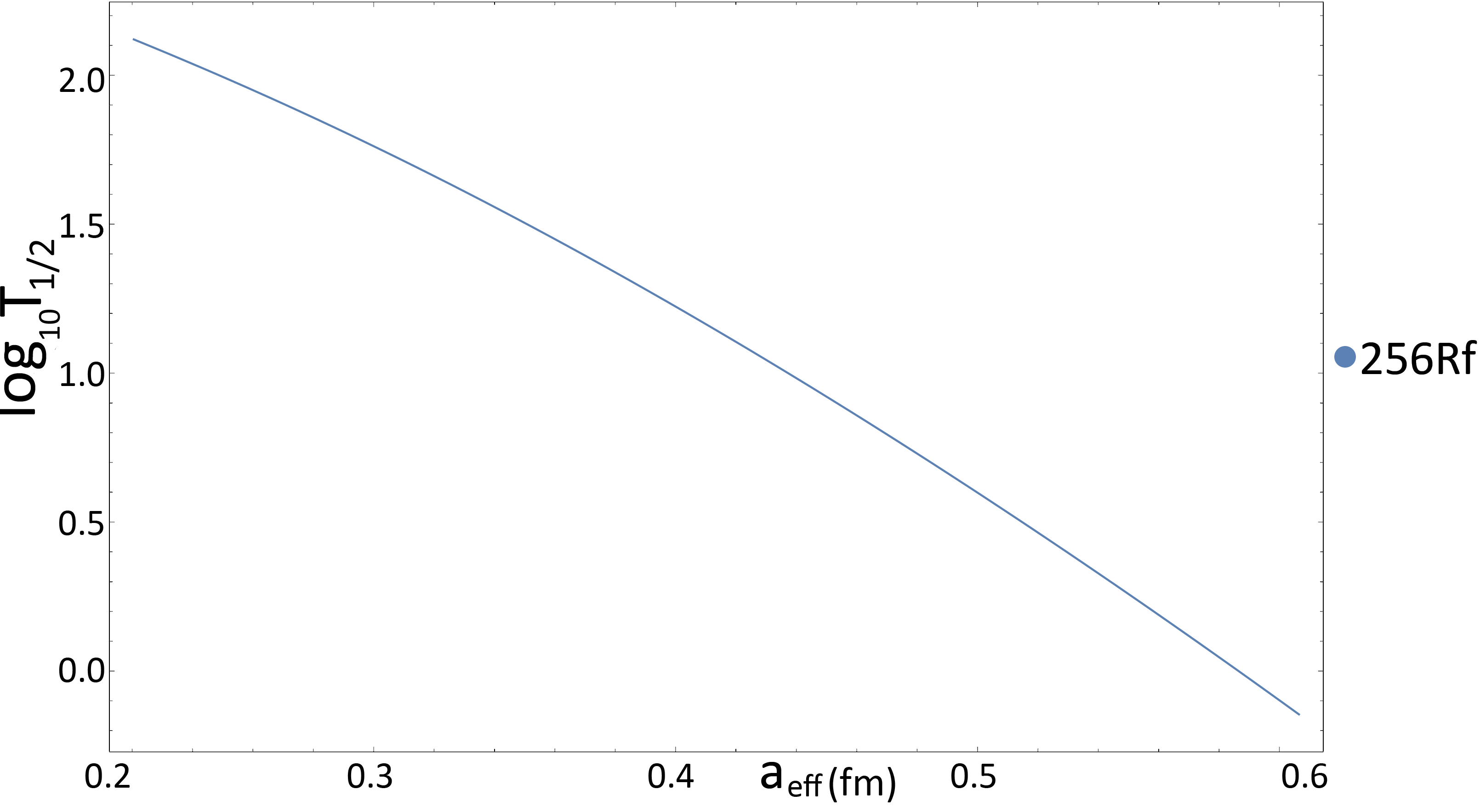}
    \caption{$\log_{10} T_{1/2}$ vs. $a_{\text{eff}}$ (fm) for 256Rf. }
    \label{fig:mesh1}
\end{figure}

\begin{figure}[htp!]
\centering
\includegraphics[width=0.5\textwidth,keepaspectratio]{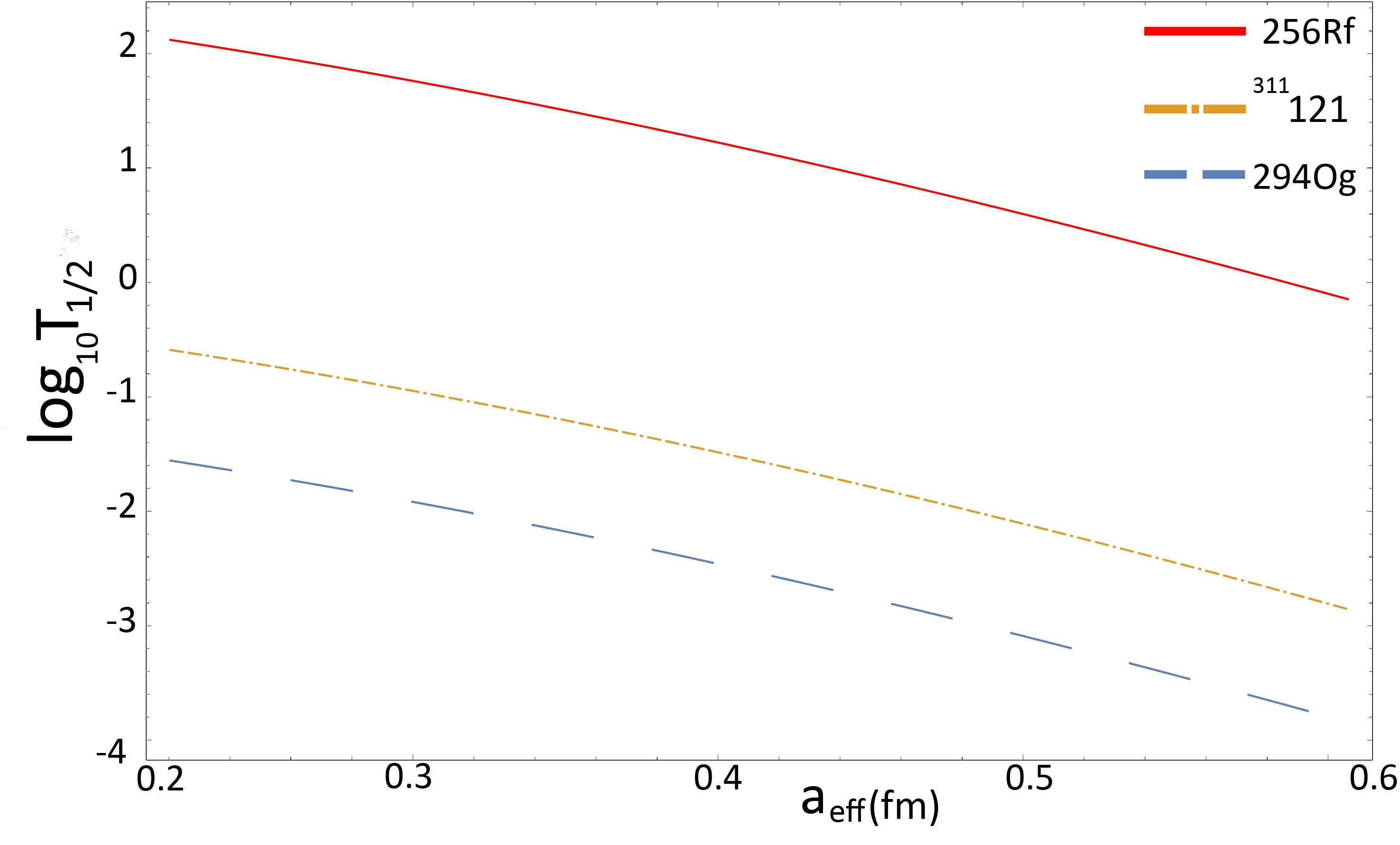}
    \caption{ $\log_{10} T_{1/2}$ vs. $a_{\text{eff}}$ (fm) for three different SHN. }
    \label{fig:mesh2}
\end{figure}

This is the motivation behind the current investigation. In this section, we shall study how half-lives of nuclei vary with $a_{\text{eff}}$. Simulation results displayed in Fig. \ref{fig:mesh1} and \ref{fig:mesh2} show variation of the logarithm of half-life with $a_{\text{eff}}$ as it's varied from 0.22 to 0.61 fm. Simulation results are interesting insofar that they intersect and diverge from the previous work which investigated the role of diffuseness \cite{roleconst}. Both results confirm the physical intuition that half-life is a monotonically decreasing function of diffuseness parameter; this is to be expected since increasing $a_{\text{eff}}$ increases the range of nuclear interaction $R$ which dominates the Coulomb force, leading to a decreased potential height and hence increased penetrability of the barrier and a decrease in half-life. 
More interesting perhaps is the different behavior exhibited by half-life when proximity potential is adopted instead of deformed WS, deformed Coulomb potentials as in the previous study \cite{roleconst}. Unlike the WS case--in which half-life varies linearly with diffuseness parameter-- the half-life here is a non-linear convex function of the effective diffuseness  $a_{\text{eff}}$. In Fig. \ref{fig:mesh1}, we can see that variation of diffuseness from 0.22 to 0.61 fm leads to change in the logarithm of half-life from 2.1 to -0.1-- two orders of magnitude change in the half-life (equivalently, $b_{\text{eff}}$ was varied from 0.4 to 1.1 fm). In addition to the significant change in half-life with small variations in diffuseness, the non-linear convex behavior implies that for systems with $a_{\text{eff}}>0.5$ fm, approximating diffuseness as 0.54 fm is problematic since half-life varies even stronger with diffuseness compared to the region $a_{\text{eff}}<0.5$ fm; to illustrate with an example, consider the case of 256Rf shown in Fig. \ref{fig:mesh1}, varying $a_{\text{eff}}$ by 0.1 fm from 0.22 to 0.32 fm induces a change of 0.457 in the logarithm of half-life, while varying $a_{\text{eff}}$ from 0.5 to 0.6 fm leads to a change in the half-life by 0.69. These two aforementioned facts-- significant change with slight diffuseness variation and convexity/non-linearity-- deem common approximations \cite{cluster,prox, constb1,constb2, constb3, constb4} such as $b_{\text{eff}}=0.99-1$ fm unacceptable since they're bound to produce large errors. Fig. \ref{fig:mesh2} helps to illustrate the universality of this behavior irrespective of the SHN under study.


\section{New semiempirical formula for $a_{\text{eff}}$} \label{sec:num4}

\begin{figure}[htp!]
\centering
\includegraphics[width=0.5\textwidth,keepaspectratio]{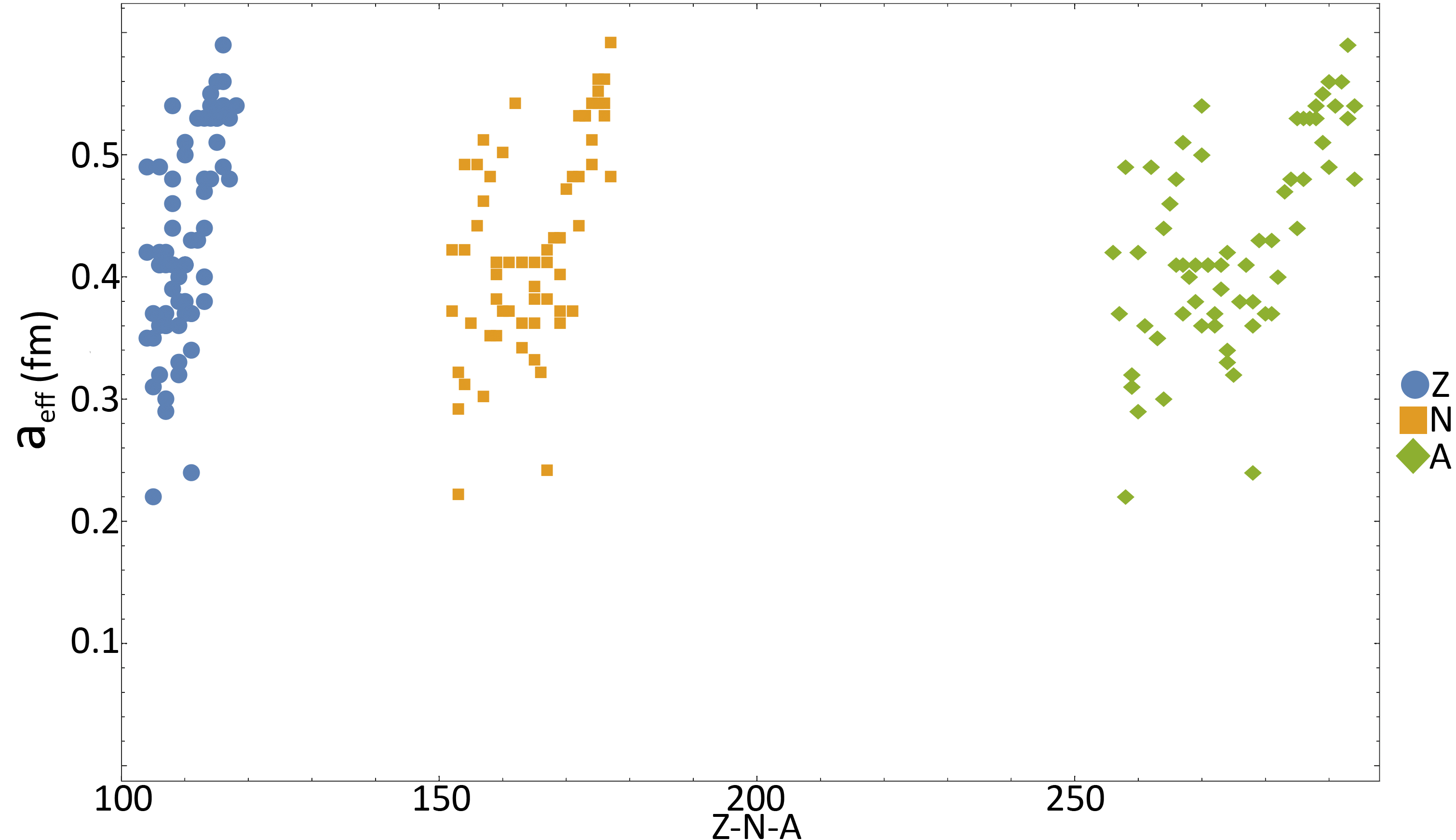}
    \caption{ $a_{\text{eff}}$ (fm) vs $Z, N$, and $A$. }
    \label{fig:mesh4}
\end{figure}

In their work, Dehghani \emph{et al.} considered deformed WS, deformed Coulomb potentials and investigated 68 SHN in the region $104\le Z \le 118$ \cite{roleconst}. For every nucleus, a systematic search was carried out looking for the optimal $a_{\text{\tiny WS}}$ value that matches the experimental half-life for that particular nucleus. The result of their work is plotted in Fig. \ref{fig:mesh3} in which diffuseness $a_{\text{\tiny WS}}$ is plotted against $Z, N$, and $A$. We see that $a_{\text{\tiny WS}}$ can be as large as 0.86 fm for the particular nucleus under study. Moreover, the general trend is an increase of $a_{\text{\tiny WS}}$ when $Z$ and $N$ (or equivalently $A$) is increased although this is not a strict rule. To map $a_{\text{\tiny WS}}$ into  $a_{\text{eff}}$, we posit the ansatz that $a_{\text{eff}}=(a_{\alpha}+a_{\text{\tiny WS}})/2$; the reasonableness and viability of this assumption will be addressed by the results in the following section. By mapping  $a_{\text{\tiny WS}}$ into $a_{\text{eff}}$, we get the plot in Fig. \ref{fig:mesh4} for effective diffuseness vs. $Z, N$, and $A$. Next, we tried fitting both  $a_{\text{\tiny WS}}$ and $a_{\text{eff}}$ vs. $Z$ and $N$. Various fitting schemes including linear and non-linear ones were considered and tried. We list the two main results of these attempts:

\begin{figure}[htp!]
\centering
\includegraphics[width=0.5\textwidth,keepaspectratio]{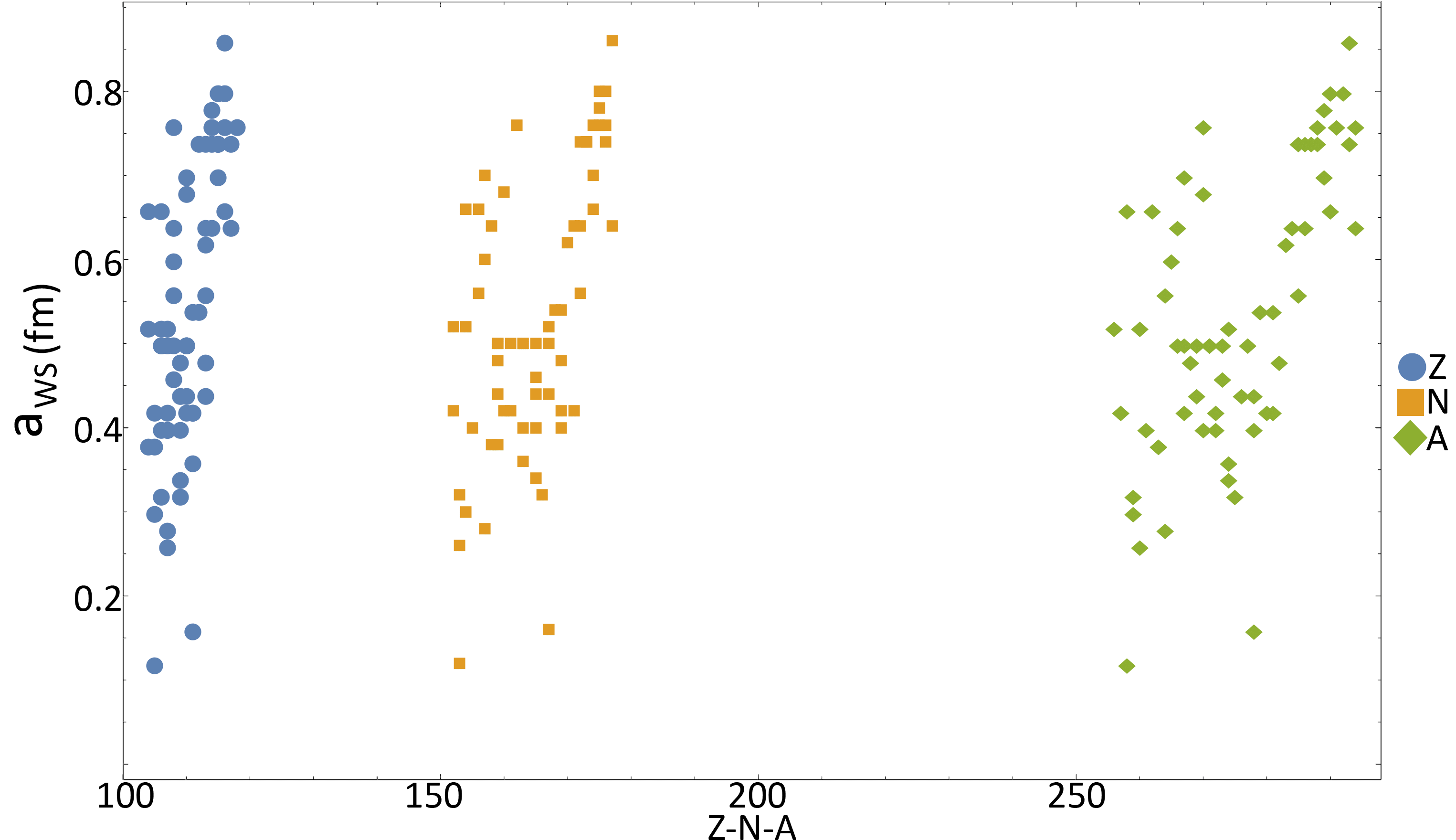}
    \caption{ $a_{\text{\tiny WS}}$ (fm) vs $Z, N$, and $A$. }
    \label{fig:mesh3}
\end{figure}

\begin{enumerate}
  \item For $a_{\text{eff}}$, a linear fit best represents the data with an rms error of 0.065 fm from true $a_{\text{eff}}$ values. The obtained semiempirical formula is given by 

\begin{equation} \label{eq:15}
a_{\text{eff}}=-1.09535+0.012063 Z +0.0019759 N
\end{equation}
  \item  No fitting model was found for $a_{\text{\tiny WS}}$ that approximate true $a_{\text{\tiny WS}}$ with a reasonable rms error comparable to the one for $a_{\text{eff}}$. In particular, linear and non-linear models produced rms error of around 0.13-0.16 fm whereas adopting constant $a_{\text{\tiny WS}}=0.54$ fm produced rms error of around 0.17 fm. 
\end{enumerate}

From comparing the figures, one can understand why it's easier to find a fit for $a_{\text{eff}}$ in which points are condensed more tightly compared to the more scattered and disperse data for $a_{\text{\tiny WS}}$. Moreover, there's a stronger $Z$ dependence in both cases compared to $N$ dependence, which is reflected in the fact that the coefficient of $Z$ is one order of magnitude larger than that of $N$. 

\section{Calculations and predictions of $\alpha$ half-lives of 218 SHN} \label{sec:num5}

In this section, we calculate and predict $\alpha$ half-lives of 218 SHN based on the theoretical framework outlined earlier and the semiempirical formula proposed. In particular, we calculate half-lives for 68 SHN for which there exists available experimental data using our proximity potential model; for these nuclei we do not use the semiempirical formula but rather compute $a_{\text{eff}}$ directly from $a_{\text{\tiny WS}}$ using the mapping outlined in the previous section; moreover, the results obtained are compared against previous work that assumed deformed WS, deformed Coulomb potentials and constant $a_{\text{\tiny WS}}=0.54$ fm for all nuclei (WS model) \cite{roleconst}. Moreover, we compare our output with the results of six of the most powerful semiempirical formulas. Regarding the 150 SHN whose half-lives to be predicted, there's no available $a_{\text{\tiny WS}}$ for them and hence we make use of our proposed semiempirical formula to compute $a_{\text{eff}}$ values. Experimental or theoretical $Q_{\alpha}$ values are taken from Refs.  \cite{improv, extended, roleconst}.

\begin{figure}[h]
\centering
\includegraphics[width=0.5\textwidth,keepaspectratio]{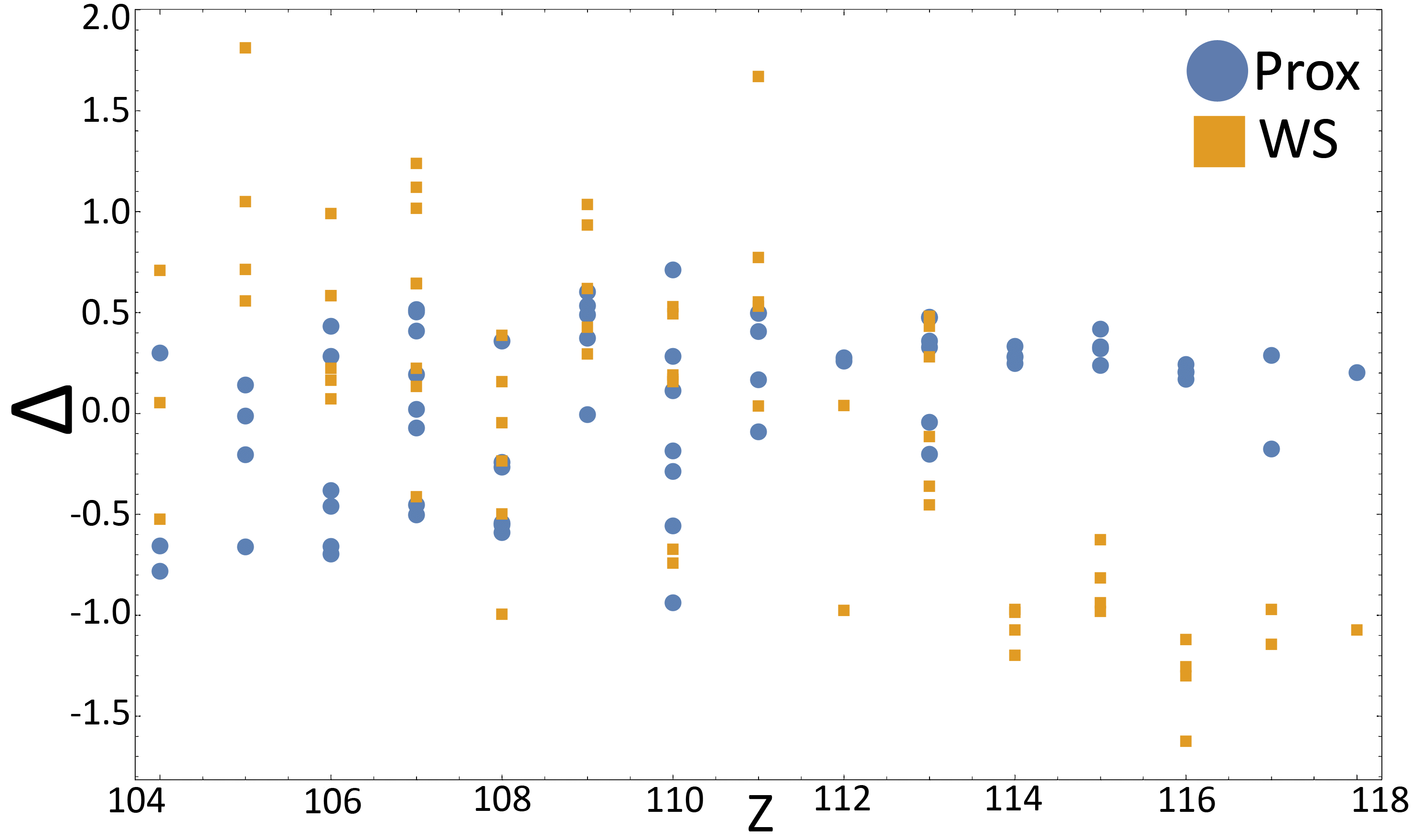}
    \caption{ $\Delta_{\text{calc}}$ and $\Delta^{\text{deformed}}_{\text{\tiny WS}}$ plotted vs. charge number $Z$ where they are dubbed prox and WS respectively.  }
    \label{fig:mesh5}
\end{figure}

The results of the calculations for the 68 SHN are tabulated in Table \ref{table:1}, where $Z$ varies from 104 to 118 and $N$ varies from 152 to 176. The table also contains the logarithm of errors for our model $\Delta_{\text{calc}}=\log_{10} T_{\text{exp}}-\log_{10} T_{\text{calc}}$ and $\Delta^{\text{deformed}}_{\text{\tiny WS}}=\log_{10} T_{\text{exp}}-\log_{10} T^{\text{deformed}}_{\text{\tiny WS}}$ for the deformed WS, deformed Coulomb model with constant diffuseness. Moreover, a plot of the errors of the two models vs. $Z$ is shown in Fig. \ref{fig:mesh5}. We note that one calculation is not included in the table, graphs or earlier fitting procedure, namely, that of 256Db with error $\Delta>1$ for both models. We worked with arbitrary precision in our calculations but rounded up numbers in the tables to the nearest 2 or 3 decimal places. Fig. \ref{fig:mesh5} is instructive to draw conclusions from; in particular, it shows our model--dubbed Prox-- performs favorably in several aspects. First, is that most of the points are clustered around the 0.5 to -0.5 band, unlike the WS model which has more disperse points and are equally likely to be found inside and outside the 0.5 band. Second, $|\Delta_{\text{calc}}|<1$ for all points in our model (except 256Db which was mentioned earlier); this is to be contrasted with the results of WS model in which there are more than 21 points exceeding the 1 to -1 band (i.e.,  $|\Delta^{\text{deformed}}_{\text{\tiny WS}}|>1$) with 3 points even exceeding 1.5 or -1.5. Statistical comparison between the performance of our model (Prox), WS model, and six semiempirical formulas (ImSahu, Sahu, Royer10, VS, SemFIS and UNIV) \cite{improv} is shown in Table \ref{table:2} in which rms, mean, mean deviation and difference between the maximum and minimum errors are shown respectively. The rms of errors is given by $\sqrt{\widebar {\delta^{2}}}$,

\begin{equation} \label{eq:16}
 \sqrt{\widebar {\delta^{2}}}= \mysqrt{\dfrac{1}{M}\mathlarger {\sum \limits_{i}^{M} \Delta_i^2}}
\end{equation}

where $M$ is the number of SHN under study. The mean of errors is simply given by

\begin{table*}[htp!]
\caption{ Experimental data $\log_{10} T_{\text{exp}}$ vs. our model Prox $\log_{10} T_{\text{calc}}$ with variable effective diffuseness $a_{\text{eff}}$ vs. $\log_{10} T^{\text{deformed}}_{\text{\tiny WS}}$  deformed Woods-Saxon, deformed Coulomb potentials with constant diffuseness $a_{\text{\tiny WS}}=0.54$ fm. Values for $Q_{\alpha}$ and $\log_{10} T^{\text{deformed}}_{\text{\tiny WS}}$ are from Ref.  \cite{roleconst}. } 
\label{table:1}
\begin{ruledtabular}
\begin{tabular}{c c c c c c c c c c c}
$Z$   & $N$   & $A$   & $Q_{\alpha}$    & $a_{\text{eff}}$ & $b_{\text{eff}}$   & $\log_{10} T_{\text{exp}}$   & $\log_{10} T_{\text{calc}}$    & $\log_{10} T^{\text{deformed}}_{\text{\tiny WS}}$  & $\Delta_{\text{calc}}$ & $\Delta^{\text{deformed}}_{\text{\tiny WS}}$\\ \hline
104 & 152 & 256 & 8.926  & 0.42   & 0.77    & 0.319  & 1.093  & 0.265 & -0.770    & 0.054  \\
104 & 154 & 258 & 9.190 & 0.49 & 0.89 &-1.035 & -0.398 & -0.512 & -0.645 & -0.523 \\
104 & 159 & 263 & 8.250   & 0.35   & 0.64 & 3.301  & 2.997  & 2.592       & 0.311    & 0.709 \\
105 & 152 & 257 & 9.206  & 0.37   & 0.67 & 0.389  & 1.039  &-0.169      & -0.649   & 0.558 \\
105 & 153 & 258 & 9.500    & 0.22   & 0.40 & 0.776  & 0.777 &-1.036      & -0.001   & 1.812 \\
105 & 154 & 259 & 9.620   & 0.31   & 0.57 &-0.292 & -0.101 &-1.341      & -0.192   & 1.049 \\
105 & 158 & 263 & 8.830   & 0.35   & 0.64 & 1.798  & 1.619  & 1.056       & 0.152    & 0.715 \\
106 & 153 & 259 & 9.804  & 0.32   & 0.58 &-0.492 & -0.043 &-1.482      & -0.449 & 0.990 \\
106 & 154 & 260 & 9.901  & 0.42   & 0.78   &-1.686 & -1.001  &-1.759      & -0.685  & 0.073           \\
106 & 155 & 261 & 9.714  & 0.36   & 0.66 &-0.638 & -0.268 &-1.222      & -0.37    & 0.584           \\
106 & 156 & 262 & 9.600    & 0.49   & 0.89 &-1.504 & -0.855 &-0.921      & -0.648   & 0.583           \\
106 & 163 & 269 & 8.700    & 0.41   & 0.75 & 2.079  & 1.785   & 1.913       & 0.294   & 0.166           \\
106 & 165 & 271 & 8.670   & 0.41   & 0.75 & 2.219  & 1.775  & 1.996       & 0.444  & 0.223           \\
107 & 153 & 260 & 10.40   & 0.29   &       0.53   &-1.459 & -0.969 &-2.698      & -0.490    & 1.239           \\
107 & 154 & 261 & 10.50   & 0.31   & 0.57    &-1.899 & -1.459 &-2.916      & -0.440    & 1.017           \\
107 & 157 & 264 & 9.960   & 0.30   & 0.55    &-0.357 & -0.390 &-1.479      & 0.033    & 1.122           \\
107 & 159 & 266 & 9.430   & 0.41   & 0.75    & 0.23   & 0.290 & 0.005 & -0.060  & 0.225           \\
107 & 160 & 267 & 9.230   & 0.37   & 0.67 & 1.23   & 1.024  & 0.643       & 0.206    & -0.413          \\
107 & 163 & 270 & 9.060   & 0.36   & 0.66    & 1.785  & 1.364  & 1.140        & 0.421    & 0.645           \\
107 & 165 & 272 & 9.310   & 0.36   & 0.66    & 1.000      & 0.474 & 0.356       & 0.526   & 0.644           \\
107 & 167 & 274 & 8.930   & 0.42   & 0.77 & 1.732  & 1.218  & 1.598       & 0.515    & 0.134           \\
108 & 156 & 264 & 10.591 & 0.44   & 0.80 &-2.796 & -2.219 &-2.750      & -0.577 & -0.046          \\
108 & 157 & 265 & 10.47  & 0.46   & 0.84 &-2.699 & -2.167 &-2.464      & -0.532  & -0.235          \\
108 & 158 & 266 & 10.346 & 0.48   & 0.87  &-2.638 & -2.097 &-2.141      & -0.54    & -0.497          \\
108 & 159 & 267 & 10.037 & 0.41   & 0.75 &-1.187 & -0.956 &-1.344      & -0.23    & 0.157           \\
108 & 162 & 270 & 9.050   & 0.54   & 0.98 & 0.556  & 0.810 & 1.55        & -0.254   & -0.994          \\
108 & 165 & 273 & 9.730   & 0.39   & 0.71 &-0.119 & -0.490 &-0.506      & 0.37     & 0.387           \\
109 & 159 & 268 & 10.67  & 0.33   & 0.60    &-1.678 & -1.684 &-2.613      & 0.006   & 0.935           \\
109 & 165 & 274 & 10.20   & 0.32   & 0.58 &-0.357 & -0.973 &-1.394      & 0.616   & 1.037           \\
109 & 166 & 275 & 10.48  & 0.38   & 0.69  &-1.699 & -2.086 &-2.127      & 0.386    & 0.428           \\
109 & 167 & 276 & 10.03  & 0.36   & 0.66 &-0.347 & -0.848 &-0.966      & 0.501    & 0.619           \\
109 & 169 & 278 & 9.580   & 0.40   & 0.73 & 0.653  & 0.106 & 0.358       & 0.547    & 0.295           \\
110 & 157 & 267 & 11.78  & 0.51   & 0.93 &-5.553 & -4.625 &-4.811      & -0.926   & -0.742          \\
110 & 159 & 269 & 11.509 & 0.38   & 0.69  &-3.747 & -3.472  &-4.242      & -0.275  & 0.495           \\
110 & 160 & 270 & 11.12  & 0.50   & 0.91     &-4.00  & -3.455 &-3.327      & -0.545   & -0.673          \\
110 & 161 & 271 & 10.899 & 0.41   & 0.75 &-2.639 & -2.465 &-2.808      & -0.174   & 0.169           \\
110 & 163 & 273 & 11.38  & 0.41   & 0.75 &-3.77  & -3.894 &-3.927      & 0.124    & 0.157           \\
110 & 167 & 277 & 10.72  & 0.41   & 0.75   &-2.222 & -2.518 &-2.413     & 0.295    & 0.191           \\
110 & 171 & 281 & 9.32   & 0.37    & 0.67    & 2.125  & 1.403  & 1.596      & 0.723    & 0.529           \\
111 & 161 & 272 & 11.197 & 0.34   & 0.62   &-2.420  & -2.338 &-3.193     & -0.080    & 0.773  \\   
111 & 163 & 274 & 11.48  & 0.24    & 0.44    &-2.194 & -2.7027 &-3.864     & 0.508    & 1.67            \\
111 & 167 & 278 & 10.85  & 0.43    & 0.78    &-2.377 & -2.557  &-2.415     & 0.180     & 0.038           \\
111 & 168 & 279 & 10.53  & 0.37    & 0.67    &-1.046 & -1.465 &-1.578     & 0.419    & 0.532           \\
111 & 169 & 280 & 9.91   & 0.37     & 0.67     & 0.663  & 0.151 & 0.111      & 0.511    & 0.552           \\
112 & 169 & 281 & 10.46  & 0.43    & 0.78    &-1.000  & -1.288 &-1.040     & 0.287    & 0.040            \\
112 & 173 & 285 & 9.32   & 0.53     & 0.96     & 1.447  & 1.175  & 2.424      & 0.2715   & -0.977          \\
113 & 165 & 278 & 11.85  & 0.38    & 0.69    &-3.62  & -3.589 &-4.053     & -0.032   & 0.433           \\
113 & 169 & 282 & 10.78  & 0.42    & 0.73  &-1.155 & -1.525 &-1.437     & 0.370     & 0.282           \\
113 & 170 & 283 & 10.48  & 0.47    & 0.86   &-1.00  & -0.812 &-0.640     & -0.189   & -0.360           \\
113 & 171 & 284 & 10.12  & 0.48    & 0.87   &-0.041 & -0.382 & 0.442      & 0.340     & 0.483           \\
113 & 172 & 285 & 10.01  & 0.44    & 0.80     & 0.623  & 0.137 & 0.738      & 0.486    & -0.115          \\
113 & 173 & 286 & 9.79   &  0.48    &   0.87  & 0.978  & 0.482 & 1.431      & 0.490     & -0.453          \\
114 & 172 & 286 & 10.35  & 0.53    & 0.96     &-0.699 & -0.994 & 0.273      & 0.294    & -0.972          \\
114 & 173 & 287 & 10.17  & 0.54    & 0.98     &-0.319 & -0.610 & 0.755      & 0.290     & -1.074          \\
114 & 174 & 288 & 10.072 & 0.55   & 1.00    & -0.180  & -0.440 & 1.018      & 0.259    & -1.198          \\
\end{tabular}
\end{ruledtabular}
\end{table*}

\begin{table*}[htp!]
\ContinuedFloat
\caption{(\textit{Continued}).} 
\label{table:1}
\begin{ruledtabular}
\begin{tabular}{c c c c c c c c c c c}

$Z$   & $N$   & $A$   & $Q_{\alpha}$    & $a_{\text{eff}}$ & $b_{\text{eff}}$   & $\log_{10} T_{\text{exp}}$   & $\log_{10} T_{\text{calc}}$    & $\log_{10} T^{\text{deformed}}_{\text{\tiny WS}}$  & $\Delta_{\text{calc}}$ & $\Delta^{\text{deformed}}_{\text{\tiny WS}}$\\ \hline

114 & 175 & 289 & 9.98   & 0.53     & 0.97     & 0.279  & -0.066 & 1.264      & 0.345    & -0.985          \\
115 & 172 & 287 & 10.76  & 0.53    & 0.97    &-1.432 & -1.682 &-0.451     & 0.249    & -0.981          \\
115 & 173 & 288 & 10.63  & 0.53    & 0.97    &-1.060  & -1.393 &-0.121     & 0.333    & -0.939          \\
115 & 174 & 289 & 10.52  & 0.51    & 0.93     &-0.658 & -1.010  & 0.156      & 0.342    & -0.814          \\
115 & 175 & 290 & 10.41  & 0.49    & 0.89     &-0.187 & -0.614  & 0.438      & 0.429    & -0.625          \\
116 & 174 & 290 & 10.99  & 0.54    & 0.98     &-1.824 & -2.047 &-0.705     & 0.220     & -1.12           \\
116 & 175 & 291 & 10.89  & 0.56    & 1.02     &-1.721 & -1.975 &-0.466     & 0.254    & -1.255          \\
116 & 176 & 292 & 10.774 & 0.59   & 1.07    &-1.745 & -1.926 &-0.121     & 0.181    & -1.624          \\
116 & 177 & 293 & 10.68  & 0.56   & 1.02    &-1.276 & -1.492 & 0.024      & 0.214    & -1.300            \\
117 & 176 & 293 & 11.18  & 0.53   & 0.97    &-1.854 & -2.156  &-0.883     & 0.301    & -0.971          \\
117 & 177 & 294 & 11.07  & 0.48   & 0.87    &-1.745 & -1.580 &-0.601     & -0.165   & -1.144          \\
118 & 176 & 294 & 11.82  & 0.54   & 0.98    &-3.161 & -3.376 &-2.089     & 0.215    & -1.072  \\
\end{tabular}
\end{ruledtabular}
\end{table*}

\begin{equation} \label{eq:17}
\widebar\delta=\dfrac{1}{M} \mathlarger{ \sum \limits_{i}^{M} \Delta_i }
\end{equation}

Mean deviation is given by

\begin{equation} \label{eq:18}
\widebar{|\delta|}=\dfrac{1}{M} \mathlarger{ \sum \limits_{i}^{M}| \Delta_i| }
\end{equation}

We note that the statistical parameters shown in Table \ref{table:2} for semiempirical formulas were obtained for 69 SHN in the region $92 \le Z \le 118$ \cite{improv}; owing to the similar sample size and same region for charge number, we conclude that it is reasonable to make a comparison between them and our model. We note that there was no available data for $\widebar \delta$ and $\Delta_{\text{max}}-\Delta_{\text{min}}$ for the semiempirical formulas.  We observe that our model Prox performs better than all models except ImSahu.  All in all, these results help to consolidate the importance of diffuseness in half-lives calculations; it shows that using accurate diffuseness parameter is more important than taking deformation into account. Moreover, we note that WS takes two parameters $\beta_2$ and $\beta_4$ (deformation parameters), while Prox takes only one, namely $a_{\text{eff}}$(If we use our new fitted formula then we take no parameters at all). In addition, it shows the reasonableness of the mapping $a_{\text{eff}}=(a_{\alpha}+a_{\text{\tiny WS}})/2$ proposed earlier. We expect that by incorporating deformation, more accurate preformation factor formula and etc. into our model, we can get even more accurate results. 

\begin{table}[htp!]
\caption{ Statistical comparison between our model Prox vs. deformed WS, deformed Coulomb model (WS) vs. ImSahu vs. Sahu vs. Royer10 vs. VS vs. SemFIS  vs. UNIV. Data for statistical parameters for semiempirical formulas are taken from Ref. \cite{improv}. } 
\label{table:2}
\begin{ruledtabular}
\begin{tabular}{c c c c c}
Model   &  $\sqrt{\widebar {\delta^{2}}}$  & $\widebar\delta$   &$\widebar{ |\delta|}$     &  $\Delta_{\text{max}}-\Delta_{\text{min}}$ \\ \hline
Prox & 0.41 & 0.064 & 0.36  & 1.65        \\
WS & 0.79 &0.01 & 0.67 & 3.44  \\
ImSahu & 0.362 &  & 0.287 &   \\
Sahu & 0.709 &  & 0.58 &   \\
Royer10 & 0.523 &  & 0.429 &   \\
VS & 0.623 &  & 0.508 &   \\
SemFIS & 0.504 &  & 0.413 &   \\
UNIV & 0.477 &  & 0.392 &   \\
\end{tabular}
\end{ruledtabular}
\end{table}

Motivated by these results, we predicted the half-lives of 150 SHN between $105\le Z \le121$ with the aid of our new semiempirical formula for $a_{\text{eff}}$. We will compare our results with seven of the current most powerful semiempirical formulas for $\alpha$ half-lives: Viola-Seaborg (VS), Royer (R), modified Brown 1 (mB1), modified Brown 2 (mB2), ImSahu, SemFIS and UNIV. Half-lives predictions made by these 7 semiempirical formulas which we shall compare our model against are taken from Refs.  \cite{extended, improv}.

Predictions of half-lives for 35 SHN in the region $105 \le Z \le 118$ are shown in Table \ref{table:3}. The comparison is made between our model Prox, Viola-Seaborg (VS), modified Brown 1 (mB1), Royer (R), and modified Brown 2 (mB2) and the average of the four semiempirical formulas \cite{extended}. The results of our model are pretty consistent with the predictions of the four formulas.

\begin{table*}[htp!]
\caption{ Our model Prox $\log_{10} T_{\text{calc}}$ vs. Viola-Seaborg $\log_{10} T_{\text{\tiny VS}}$ vs. modified Brown formula 1 $\log_{10} T_{\text{\tiny mB1}}$ vs. Royer $\log_{10} T_{\text{\tiny R}}$ vs. modified Brown formula 2 $\log_{10} T_{\text{\tiny mB2}}$ vs. average of the four semiempirical formulas  $\log_{10} T_{\text{avg}}$ and error  $\Delta_{\text{avg}}$ between our model and the average. $Q_{\alpha}$ values and half-lives for the semiempirical formulas are from Ref.  \cite{extended}.} 
\label{table:3}
\begin{ruledtabular}
\begin{tabular}{c  c  c  c  c  c  c  c  c  c  c}
$Z$   & $N$   & $A$   & $Q_{\alpha}$     &  $\log_{10} T_{\text{calc}}$    &  $\log_{10} T_{\text{\tiny VS}}$    & $\log_{10} T_{\text{\tiny mB1}}$  & $\log_{10} T_{\text{\tiny R}}$     &  $\log_{10} T_{\text{\tiny mB2}}$   & $\log_{10} T_{\text{avg}}$   & $\Delta_{\text{avg}}$ \\ \hline
105 & 159 & 264 & 8.95  & 1.096  & 1.179  & 1.181  & 1.078  & 1.118  & 1.139  & 0.043   \\
106 & 161 & 267 & 9.12  & 0.790 & 0.641  & 0.622  & 0.666  & 0.685  & 0.654  & -0.013   \\
106 & 167 & 273 & 8.20   & 3.493  & 3.239  & 3.247  & 3.367  & 3.439  & 3.323  & -0.170    \\
108 & 153 & 261 & 10.97 & -2.402 &-3.164 &-3.197 &-3.228 &-3.335 &-3.231 & -0.830    \\
108 & 164 & 272 & 9.60   & -0.123  &-0.558 &-0.613 &-0.597 &-0.562 &-0.583 & -0.460  \\
108 & 169 & 277 & 8.85  & 1.910  & 1.859  & 1.822  & 1.88   & 1.928  & 1.872  & -0.036   \\
110 & 165 & 275 & 10.38 & -1.627  &-1.483 &-1.503 &-1.590 &-1.573 &-1.537 & 0.090   \\
110 & 166 & 276 & 10.23 & -1.310 &-1.603 &-1.645 &-1.749 &-1.714 &-1.678 & -0.369  \\
110 & 177 & 278 & 9.94  & -0.651 &-0.919 &-0.970 &-0.952 &-0.907 &-0.937 & -0.287   \\
111 & 165 & 276 & 10.44 & -1.434 &-1.067 &-1.072 &-0.983 &-1.001 &-1.031 & 0.403   \\ 
112 & 160 & 272 & 11.20  & -2.402  &-3.271 &-3.257 &-3.490 &-3.563 &-3.395 & -0.993   \\
112 & 161 & 273 & 11.06 & -2.199 &-2.543 &-2.517 &-2.565 &-2.650 &-2.569 & -0.370    \\
112 & 162 & 274 & 10.92 & -1.980 &-2.703 &-2.703 &-2.838 &-2.900 &-2.786 & -0.806   \\
112 & 163 & 275 & 10.79 & -1.772 &-1.961 &-1.949 &-1.980 &-2.054 &-1.986 & -0.214   \\
112 & 164 & 276 & 10.65 & -1.525 &-2.107 &-2.120 &-2.149 &-2.203 &-2.145 & -0.620    \\
112 & 165 & 278 & 10.37 & -0.985  &-1.478 &-1.507 &-1.421 &-1.468 &-1.469 & -0.484  \\
112 & 166 & 279 & 10.23 & -0.692 &-0.704 &-0.721 &-0.709 &-0.768 &-0.726 & -0.034 \\
113 & 167 & 280 & 10.69 & -1.551 &-1.192 &-1.170 &-1.085 &-1.126 &-1.143 & 0.416    \\

114 & 164 & 278 & 11.55 & -2.990 &-3.594 &-3.529 &-3.778 &-3.814 &-3.679 & -0.691\\
114 & 165 & 279 & 11.41 & -2.780 &-2.876 &-2.801 &-2.923 &-2.961 &-2.890 & -0.111   \\
114 & 166 & 280 & 11.28 & -2.577 &-3.047 &-3.000 &-3.151 &-3.180 &-3.095 & -0.518   \\
114 & 167 & 281 & 11.14 & -2.339 &-2.315 &-2.260 &-2.362 &-2.394 &-2.333 & 0.006    \\
114 & 168 & 282 & 11.00    & -2.086 &-2.472 &-2.445 &-2.491 &-2.515 &-2.481 & -0.390    \\
114 & 169 & 283 & 10.87 & -1.845 &-1.726 &-1.691 &-1.770 &-1.798 &-1.746 & 0.098    \\
114 & 170 & 284 & 10.73 & -1.563 &-1.868 &-1.862 &-1.795 &-1.816 &-1.835 & -0.272   \\
114 & 171 & 285 & 10.59 & -1.267 &-1.107 &-1.093 &-1.146 &-1.171 &-1.129 & 0.137    \\
115 & 171 & 286 & 11.04 & -2.093 &-1.566 &-1.500 &-1.447 &-1.455 &-1.492 & 0.600      \\
115 & 176 & 291 & 10.35 & -0.549 &-0.207 &-0.207 &-0.224 &-0.216 &-0.214 & 0.335    \\
116 & 171 & 287 & 11.50  & -2.886 &-2.664 &-2.554 &-2.731 &-2.715 &-2.666 & 0.220     \\
116 & 172 & 288 & 11.36 & -2.628 &-2.832 &-2.753 &-2.819 &-2.808 &-2.803 & 0.175    \\
116 & 173 & 289 & 11.22 & -2.356 &-2.097 &-2.012 &-2.164 &-2.148 &-2.105 & 0.250     \\
117 & 175 & 292 & 11.4  & -2.607 &-1.934 &-1.812 &-1.798 &-1.768 &-1.828 & 0.772    \\
118 & 175 & 293 & 11.85 & -3.367 &-3.010 &-2.833 &-3.089 &-3.020 &-2.988 & 0.378    \\
118 & 177 & 295 & 11.58 & -2.847 &-2.463 &-2.316 &-2.545 &-2.479 &-2.451 & 0.395     \\

\end{tabular}
\end{ruledtabular}
\end{table*}

\begin{table*}[htp!]
\caption{ Statistical comparison showing how consistent our model Prox is on one hand with VS, mB1, Royer, and mB2 on the other. } 
\label{table:4}
\begin{ruledtabular}
\begin{tabular}{ c c c c c}
Model   &  $\sqrt{\widebar {\delta^{2}}}$  & $\widebar\delta$   &$\widebar{ |\delta|}$     &  $\Delta_{\text{max}}-\Delta_{\text{min}}$ \\ \hline
Prox & 0.43 & -0.09 & 0.34  & 1.77        \\
\end{tabular}
\end{ruledtabular}
\end{table*}

\FloatBarrier

The most relevant statistical parameters are shown in Table \ref{table:4}. We can see that the values of the parameters are consistent with the ones in Table \ref{table:2} reassuring us further to the utility and viability of our fit. 

Next, we predict the half-lives of 115 SHN in the $118\le Z \le121$ region and compare our results to that of ImSahu, SemFIS and UNIV semiempirical formulas \cite{improv}. The predictions are compiled in Table \ref{table:5}. To perform meaningful statistical analysis, we considered logarithm of half-lives values in which the three semiempirical formulas converge within 10$\%$; hence, we define $\epsilon$ as \cite{extended},

\begin{equation} \label{eq:18}
\epsilon= \dfrac{\max(|\log_{10} T_i|)-\min(|\log_{10} T_j|)}{\max(|\log_{10} T_i|)}
 \end{equation}

where the indices $i$ and $j$ can be 1, 2, and 3 representing ImSahu, SemFIS, and UNIV, respectively. After looking for SHN with $\epsilon<0.1$, we took the average of the three formulas $\log_{10} T_{\text{avg}}=(\log_{10} T_{\text{\tiny ImSahu}}+\log_{10} T_{\text{\tiny SemFIS}} +\log_{10} T_{\text{\tiny UNIV}})/3$ then defined 
$\Delta_{\text{avg}}=\log_{10} T_{\text{avg}}-\log_{10} T_{\text{calc}}$ to measure consistency of our results with that of the other formulas. The result of this analysis is shown in Table \ref{table:6}. The rms error is 0.526 which is still considered acceptable with current standards.

\begin{table*}[htp!]
\caption{ Half-lives calculations for our model Prox $\log_{10} T_{\text{calc}}$  vs. ImSahu  $\log_{10} T_{\text{\tiny ImSahu}}$ vs. SemFIS  $\log_{10} T_{\text{\tiny SemFIS}}$ vs. UNIV $\log_{10} T_{\text{\tiny UNIV}}$. $Q_{\alpha}$ values and half-lives for the semiempirical formulas are from Ref.  \cite{improv}.} 
\label{table:5}
\begin{ruledtabular}
\begin{tabular}{c c c c c c c c}
$Z$   & $N$   & $A$   & $Q_{\alpha}$     &  $\log_{10} T_{\text{calc}}$    &  $\log_{10} T_{\text{\tiny ImSahu}}$   & $\log_{10} T_{\text{\tiny SemFIS}}$  & $\log_{10} T_{\text{\tiny UNIV}}$ \\ \hline
118 & 160 & 278 & 13.89 & -5.693 & -6.69  & -7.32  & -7.26 \\
118 & 161 & 279 & 13.78 & -5.668 & -8.25  & -6.44  & -6.4  \\
118 & 162 & 280 & 13.71 & -5.705 & -6.46  & -6.90  & -6.99 \\
118 & 163 & 281 & 13.76 & -5.948 & -8.14  & -6.31  & -6.40 \\
118 & 164 & 282 & 13.49 & -5.606 & -6.15  & -6.42  & -6.64 \\
118 & 165 & 283 & 13.33 & -5.447 & -7.32  & -5.46  & -5.68 \\
118 & 166 & 284 & 13.23 & -5.389 & -5.75  & -5.87  & -6.21 \\
118 & 167 & 285 & 13.07 & -5.207 & -6.77  & -4.91  & -5.24 \\
118 & 168 & 286 & 12.92 & -5.029 & -5.25  & -5.2   & -5.67 \\
118 & 169 & 287 & 12.8  & -4.900 & -6.19  & -4.35  & -4.76 \\
118 & 170 & 288 & 12.62 & -4.636 & -4.75  & -4.62  & -5.12 \\
118 & 171 & 289 & 12.59 & -4.668 & -5.71  & -3.93  & -4.39 \\
118 & 172 & 290 & 12.6  & -4.775 & -4.77  & -4.59  & -5.11 \\
118 & 173 & 291 & 12.42 & -4.479 & -5.30  & -3.60  & -4.08 \\
118 & 174 & 292 & 12.24 & -4.167 & -4.12  & -3.87  & -4.42 \\
118 & 175 & 293 & 12.24 & -4.231 & -4.85  & -3.28  & -3.74 \\
118 & 176 & 294 & 11.82 & -3.355  & -3.30  & -3.02  & -3.53 \\
118 & 177 & 295 & 11.9  & -3.585 & -4.06  & -2.62  & -3.05 \\
118 & 178 & 296 & 11.75 & 3.285  & -3.21  & -2.97  & -3.43 \\
118 & 179 & 297 & 12.1  & -4.106 & -4.41  & -3.19  & -3.51 \\
118 & 180 & 298 & 12.18 & -4.307 & -4.19  & -4.06  & -4.39 \\
118 & 181 & 299 & 12.05 & -4.041 & -4.22  & -3.23  & -3.44 \\
118 & 182 & 300 & 11.96 & -3.852 & -3.79  & -3.76  & -3.95 \\
118 & 183 & 301 & 12.02 & -3.990 & -4.08  & -3.36  & -3.41 \\
118 & 184 & 302 & 12.04 & -4.030 & -4.03  & -4.13  & -4.16 \\
118 & 185 & 303 & 12.6  & -5.213 & -5.18  & -4.77  & -4.62 \\

118 & 186 & 304 & 13.12 & -6.227 & -6.20  & -6.49  & -6.32 \\
118 & 187 & 305 & 12.91 & -5.792 & -5.69  & -5.59  & -5.25 \\
118 & 188 & 306 & 12.48 & -4.885 & -5.05  & -5.52  & -5.12 \\
118 & 189 & 307 & 11.92 & -3.631 & -3.63  & -3.90  & -3.28 \\
118 & 190 & 308 & 12.2  & -4.200 & -2.34  & -3.10  & -2.35 \\
119 & 161 & 280 & 14.3  & -6.224 & -7.32  & -6.51  & -6.30 \\
119 & 162 & 281 & 14.16 & -6.154 & -7.26  & -6.97  & -6.96 \\
119 & 163 & 282 & 14    & -6.038 & -6.90  & -5.90  & -5.84 \\
119 & 164 & 283 & 13.76 & -5.766 & -6.60  & -6.19  & -6.32 \\
119 & 165 & 284 & 13.57 & -5.566 & -6.24  & -5.05  & -5.14 \\
119 & 166 & 285 & 13.61 & -5.777 &-6.34  &-5.84  &-6.10 \\
119 & 167 & 286 & 13.43 & -5.575 &-6.06  &-4.73  &-4.92 \\
119 & 168 & 287 & 13.28 & -5.415 &-5.76  &-5.17  &-5.54 \\
119 & 169 & 288 & 13.23 & -5.435 &-5.77  &-4.31  &-4.59 \\
119 & 170 & 289 & 13.16 & -5.408 &-5.54  &-4.91  &-5.35 \\
119 & 171 & 290 & 13.07 & -5.334 &-5.54  &-3.98  &-4.33 \\
119 & 172 & 291 & 13.05 & -5.388 &-5.02  &-3.35  &-3.74 \\
119 & 173 & 292 & 12.9  & -5.176 &-4.78  &-4.20  &-4.69 \\
119 & 174 & 293 & 12.72 & -4.890 &-4.59  &-2.19  &-3.28\\ 
119 & 175 & 294 & 12.73 & -4.891 &-4.90  &-3.33  &-3.52 \\

119 & 185 & 304 & 12.93 & -5.677 &-5.71  &-4.39  &-4.28 \\
119 & 186 & 305 & 13.42 & -6.609 &-5.97  &-6.15  &-6.07 \\
119 & 187 & 306 & 13.2  & -6.172 &-6.27  &-5.10  &-4.82 \\
119 & 188 & 307 & 12.78 & -5.316 &-4.80  &-5.18  &-4.91 \\
119 & 189 & 308 & 12.06 & -3.752 &-4.11  &-3.11  &-2.59 \\
119 & 190 & 309 & 11.37 & -2.097 &-1.81  &-2.49  &-1.92 \\
120 & 163 & 283 & 14.31 & -6.238 &-9.02  &-6.87  &-6.83 \\
120 & 164 & 284 & 13.99 & -5.843 &-6.45  &-6.91  &-7.01 \\

\end{tabular}
\end{ruledtabular}
\end{table*}

\begin{table*}[htp!]
\ContinuedFloat
\caption{ (\textit{Continued}).} 
\label{table:5}
\begin{ruledtabular}
\begin{tabular}{c c c c c c c c}
$Z$   & $N$   & $A$   & $Q_{\alpha}$     &  $\log_{10} T_{\text{calc}}$    &  $\log_{10} T_{\text{\tiny ImSahu}}$   & $\log_{10} T_{\text{\tiny SemFIS}}$  & $\log_{10} T_{\text{\tiny UNIV}}$ \\ \hline
120 & 165 & 285 & 13.89 & -5.818 &-8.25  &-6.05  &-6.17 \\
120 & 166 & 286 & 14.03 & -6.211 &-6.58  &-6.88  &-7.11 \\
120 & 167 & 287 & 13.85 & -6.029 &-8.11  &-5.90  &-6.13 \\
120 & 168 & 288 & 13.73 & -5.942 &-6.15  &-6.28  &-6.63 \\
120 & 169 & 289 & 13.71 & -6.028 &-7.79  &-5.59  &-5.92 \\
120 & 170 & 290 & 13.7  & -6.124 &-6.16  &-6.18  &-6.61 \\
120 & 171 & 291 & 13.51 & -5.880 &-7.36  &-5.19  &-5.60 \\
120 & 172 & 292 & 13.47 & -5.905 &-5.82  &-5.74  &-6.24 \\
120 & 173 & 293 & 13.4  & -5.865 &-7.09  &-4.98  &-5.43 \\
120 & 174 & 294 & 13.24 & -5.644 &-5.47  &-5.31  &-5.85 \\
120 & 175 & 295 & 13.27 & -5.779 &-6.77  &-4.76  &-5.23 \\

120 & 176 & 296 & 13.34 & -5.983 &-5.72  &-5.54  &-6.06 \\
120 & 177 & 297 & 13.14 & -5.659 &-6.46  &-4.55  &-5.02 \\
120 & 178 & 298 & 13.01 & -5.457 &-5.18  &-4.97  &-5.48 \\
120 & 179 & 299 & 13.26 & -5.994 &-6.60  &-4.87  &-5.27 \\
120 & 180 & 300 & 13.32 & -6.149 &-5.81  &-5.66  &-6.09 \\
120 & 181 & 301 & 13.06 & -5.676 &-6.15  &-4.59  &-4.93 \\
120 & 182 & 302 & 12.89 & -5.361 &-5.08  &-4.95  &-5.31 \\
120 & 183 & 303 & 12.81 & -5.216 &-5.60  &-4.23  &-4.48 \\
120 & 184 & 304 & 12.76 & -5.123 &-4.89  &-4.84  &-5.09 \\
120 & 185 & 305 & 13.28 & -6.158 &-6.40  &-5.31  &-5.40 \\
120 & 186 & 306 & 13.79 & -7.102 &-6.82  &-6.95  &-7.01 \\
120 & 187 & 307 & 13.52 & -6.594 &-6.74  &-5.94  &-5.86 \\
120 & 188 & 308 & 12.97 & -5.512 &-5.42  &-5.65  &-5.56 \\
120 & 189 & 309 & 12.16 & -3.779 &-4.05  &-3.51  &-3.25 \\
120 & 190 & 310 & 11.5  & -2.211 &-2.41  &-2.82  &-2.48 \\
121 & 165 & 286 & 14.34 & -6.279 &-7.31  &-7.26  &-5.96 \\
121 & 166 & 287 & 14.53 & -6.752 &-7.56  &-7.53  &-7.15 \\
121 & 167 & 288 & 14.46 & -6.778 &-7.56  &-7.37  &-6.18 \\
121 & 168 & 289 & 14.4  & -6.813 &-7.36  &-7.23  &-6.97 \\
121 & 169 & 290 & 14.42 & -6.976 &-7.56  &-7.24  &-6.15 \\
121 & 170 & 291 & 14.4  & -7.064 &-7.35  &-7.18  &-7.00 \\
121 & 171 & 292 & 14.31 & -7.024 &-7.45  &-7.01  &-6.00 \\
121 & 172 & 293 & 14.1  & -6.766 &-6.86  &-6.63  &-6.54 \\
121 & 173 & 294 & 14.1  & -6.865 &-7.17  &-6.63  &-5.69 \\
121 & 174 & 295 & 13.98 & -6.744 &-6.66  &-6.42  &-6.37 \\
121 & 175 & 296 & 14.01 & -6.881 &-7.08  &-6.48  &-5.57 \\
121 & 176 & 297 & 14.12 & -7.152 &-6.89  &-6.68  &-6.64 \\
121 & 177 & 298 & 13.89 & -6.812 &-6.94  &-6.30  &-5.39 \\
121 & 178 & 299 & 13.65 & -6.435 &-6.09  &-5.88  &-5.86 \\
121 & 179 & 300 & 13.81 & -6.783 &-6.87  &-6.21  &-5.29 \\
121 & 180 & 301 & 13.82 & -6.847 &-6.38  &-6.27  &-6.19 \\
121 & 181 & 302 & 13.49 & -6.275 &-6.38  &-5.71  &-4.75 \\
121 & 182 & 303 & 13.31 & -5.962 &-5.48  &-5.42  &-5.31 \\
121 & 183 & 304 & 13.28 & -5.928 &-6.06  &-5.43  &-4.40 \\
121 & 184 & 305 & 13.27 & -5.925 &-5.40  &-5.48  &-5.27 \\
121 & 185 & 306 & 13.81 & -6.948 &-7.06  &-6.55  &-5.38 \\
121 & 186 & 307 & 14.34 & -7.883 &-7.23  &-7.54  &-7.15 \\
121 & 187 & 308 & 14.07 & -7.409 &-7.55  &-7.17  &-5.85 \\
121 & 188 & 309 & 13.26 & -5.896 &-5.38  &-5.81  &-5.31 \\
121 & 189 & 310 & 12.46 & -4.238 &-4.66  &-4.34  &-2.89 \\
121 & 190 & 311 & 11.81 & -2.747 &-2.44  &-3.06  &-2.37 \\

\end{tabular}
\end{ruledtabular}
\end{table*}

\begin{table*}[htp!]
\caption{ Measure of consistency of our model Prox with that of ImSahu, SemFIS, and UNIV.  } 
\label{table:6}
\begin{ruledtabular}
\begin{tabular}{  c c c c c}
Model   &  $\sqrt{\widebar {\delta^{2}}}$  & $\widebar\delta$   &$\widebar{ |\delta|}$     &  $\Delta_{\text{max}}-\Delta_{\text{min}}$   \\ \hline
Prox & 0.526 & -0.013 & 0.4217  & 2.46        \\
\end{tabular}
\end{ruledtabular}
\end{table*}

\FloatBarrier

\section{Discussion and conclusions} \label{sec:num6}

The present analysis and results leave us with several conclusions and remarks:

\begin{itemize}
  \item The effect of diffuseness can no longer be ignored by adopting it as a constant for all SHN; even very slight variations as small as 0.1 fm can induce an error in half-life as large as 0.69; this is especially true for systems which we expect to have large diffuseness; the effect is especially pronounced when proximity potential is adopted in which we find non-linear (convex) variation of the logarithm of half-life with diffuseness; we conclude that diffuseness is a great bottleneck acting as a limiting factor against accurate half-lives calculations.  
  \item Present calculations and comparisons show that taking true values of diffuseness parameter into account is more critical than deformation effects; this was shown when we compared our model with WS model.
  \item  The use of the mapping $a_{\text{eff}}=(a_{\alpha}+a_{\text{\tiny WS}})/2$ produced results for 68 SHN that are in great agreement with available experimental data; it outperformed all models except the ImSahu model. This shows the viability and utility of this mapping, which can be adopted for future use. 
 \item  Our predictions of $\alpha$ half-lives of 150 SHN using our newly proposed fitted formula is in good agreement with other semiempirical formulas; this is especially true for the region $105 \le Z \le 118$ which was the region of the original fit. By using a bigger sample size over a more extended region, a semiempirical formula better than the one at hand can be produced to be used in half-lives calculations.
 \item  For our predictions, we did not incorporate centripetal contribution $V_l(r)$ due to the lack of information about angular momentum of the systems under study. Incorporating these data will improve the accuracy of the results.
\item  Present investigations and calculations show that $a_{\text{eff}}$ is best represented as a linear function of charge and neutron numbers. 
 \item Present calculations corroborate the viability of the Prox model; we expect that by taking further effects into account including deformation, more accurate parity-dependent preformation factor and etc., the errors will be reduced even further. 
\item A potentially fruitful research direction is to investigate the role of diffuseness in half-life using other models for nuclear potential.
\item We stress that future research should invest an appreciable amount of energy in calculating and extracting accurate diffuseness parameters values.
  
\end{itemize}

\providecommand{\noopsort}[1]{}\providecommand{\singleletter}[1]{#1}%

\end{document}
%